\documentclass[pdflatex,sn-mathphys-num,iicol]{sn-jnl}

\usepackage{graphicx}%
\usepackage{multirow}%
\usepackage{amsmath,amssymb,amsfonts}%
\usepackage{amsthm}%
\usepackage{mathrsfs}%
\usepackage[title]{appendix}%
\usepackage{xcolor}%
\usepackage{textcomp}%
\usepackage{manyfoot}%
\usepackage{booktabs}%
\usepackage{algorithm}%
\usepackage{algorithmicx}%
\usepackage{algpseudocode}%
\usepackage{listings}%

\begin{document}

\title[The Aura in the Machine]{The Aura in the Machine: Genealogy and the Status of the Work of Art in the Generative Era}

\author*[1]{\fnm{Giorgio} \sur{Presti}}\email{giorgio.presti@unimi.it}

\affil*[1]{\orgdiv{Department of Computer Science}, \orgname{Università degli Studi di Milano}, \orgaddress{\street{Via G. Celoria}, \city{Milan}, \postcode{20133}, \country{Italy}}}

\abstract{This paper frames Generative Artificial Intelligence (AI) not as an unprecedented technological rupture, but as an industrial-scale manifestation of a deeply rooted historical process. Through a genealogy of generative arts, it shows how AI's questions on authorship and creativity have precise historical precedents.

A taxonomy of generative systems is proposed across three functional categories (\textit{medium}, \textit{artwork}, \textit{instrument}), the attribution of which is editorial rather than ontological.

From individual cognitive atrophy to \textit{Model Collapse}, the systemic risks of creative automation are identified; \textit{environmental enrichment} is proposed as an antidote.

The role of the artist undergoes a radical metamorphosis: from \textit{craftsman of the object} to \textit{entropic agent}, \textit{systems designer}, \textit{explorer}, and \textit{negentropic curator}. This pipeline-based taxonomy rests on a specific premise: the algorithmic system remains medium, instrument, or artwork, while creative agency resides in the humans distributed along it.

\textit{Algorithmic Repetition} is introduced as the aesthetic degeneration of aligned generative systems; the Benjaminian aura does not dissolve in the generative era but condenses upon the productive system. \textit{Manifestation} is proposed as a third ontological status for generative works, transcending the dichotomy between \textit{original} and \textit{copy}.

To support the proposed theses, two complementary aspects are examined: the radicalization of distributed authorship; and the reevaluation of older generative models, whose instability constitutes an aesthetic degree of freedom lost by recent ones.}


\maketitle

\section{Preface}

Navigating the space of art and Generative AI requires overcoming what Charles Percy Snow called the ``Two Cultures'' problem \cite{snow1959}: the historical disconnect between humanists and technologists. Generative AI is precisely the point where these two cultures are forced to intersect: one cannot effectively use a tool that they do not understand, nor can one understand the tool without the cultural background necessary to interpret its implications.

This contribution therefore addresses the issue without any discontinuity between technical and humanistic aspects, aiming to provide an integrated framework that is useful both to those working in art and those in technology, and especially to those operating at the intersection of these two cultures.

While recent literature applies the Benjaminian framework to Generative AI \cite{fernandez2023ai, park2024work, ambrosini2025} and proposes taxonomies of distributed authorship \cite{goodfellow2024distributed}, and analytical philosophy addresses the ontological status of generative works in terms of functional continuity \cite{norouzi2026canvas}, this contribution distinguishes itself by: a historical genealogy rooting current questions in pre-digital artistic practices; a taxonomy of roles mapped onto the Generative AI technical pipeline; and the concept of \textit{manifestation} as a third ontological status that transcends (rather than redefines) the original/copy dichotomy.

Finally, a necessary terminological clarification: When this paper attributes creativity or authorship to human beings and denies them (at present) to algorithmic systems, the criterion is not ideological but epistemic: the capacity for embodied experience of the world is considered here a fundamental requirement for the emergence of creativity \cite{Rohrmeier-2022, damasio1999, varela1991}. This capacity is, in principle, extendable to other biological agents and, hypothetically, to future artificial systems endowed with a body or capable of experiencing emotions. The limitations attributed to AI herein are therefore contingent upon current architectures, rather than absolute properties of any possible artificial system. In instances where the text uses ``human'' as an adjective, it should be read in this broader sense.

\section{Data, Rules, and Knowledge}

To understand the dynamics of Generative AI and its capacity to produce results of such complexity, it is necessary to take a step back and analyze the logical and computational elements underpinning it. In fact, AI systems do not operate through intuition or by virtue of an unfathomable principle, but rather rest on extremely concrete foundations: they process immense masses of informational fragments which, interacting on macroscopic scales, give rise to unexpected emergent phenomena \cite{anderson1972}. Before exploring the mechanisms of neural networks or their artistic applications, it is essential to understand how, starting from an inert mass of data, one can arrive at an active representation of knowledge; one so complex that it succeeds in mimicking certain traits of biological creativity.

In the field of computer science, \textit{knowledge} (information usable by a machine) can be structured in various ways. One method is \textit{trees}, such as the encyclopedia of Diderot and d'Alembert \cite{diderot1751}, which first attempted to represent all human knowledge as a tree. A more refined, ontologically robust alternative is graph-based representations like \textit{knowledge graphs}: entities linked by relationships that can be traversed to retrieve desired information. Knowledge may also be represented specifically through a \textit{set of rules}, an implicit representation similar to flowcharts or software where knowledge is crystallized into rules executed to produce responses. Finally, it can be represented via the \textit{parameters of a machine learning system} (such as the parameters of a regression function or the weights of a neural network).

Trees and graphs are inert: they are organized information that does not represent anything executable. The rules and parameters of a machine learning system, by contrast, are active, executable representations. Specifically, the distinction between these latter two models (the one based on sets of rules and the one based on machine learning systems) represents two very different approaches to software development. In the first case, there is \textit{traditional} software, in which the developer analyzes data and defines an algorithm that makes decisions. In the second case, through machine learning, the process works in reverse: data and decisions are fed into a system that, via statistical techniques, outputs rules (to be subsequently applied to new data to generate new decisions).

These principles, rarely brought into focus outside a technical context, form the foundation of this discussion, in which the inquiry into ``\textit{how}, \textit{where}, and \textit{what} knowledge is encoded'' guides reflections on authorship, the role of the artist, and the nature of the work.

\section{Programming as a Creative Act and Philosophical Exercise}
\label{sec:coding}

To frame all of this within the context of the relationship between art and generative systems, it is necessary to revisit at least a core set of concepts from art history and philosophy through the modern lenses of computer science and AI (for a more in-depth historical overview, see future publications).

The way traditional software is written is rooted in a pre-digital culture.
In an evocative metaphor, Matt Butcher \cite{butcher2012} draws a parallel to the paradigms proposed by Plato and Aristotle. Plato defines earthly entities as imperfect instances of immutable, abstract concepts, contrasting with the Aristotelian view that there exists only a material, earthly reality shaped by specific processes. Reinterpreting these two philosophers today, one can associate Platonic thought with \textit{object-oriented programming} (abstract classes, instantiated as objects), whereas Aristotelian thought can be seen as an archetype of \textit{functional programming} (data transformed by functions).

If it is true that philosophy consists of a creative act aimed at modeling reality, then programming must also inherit these characteristics \cite{sep-computer-science}: Programming consists of creating an ontology, a \textit{metaphysical system} regarding a specific problem. Programming is the creation of new worlds, leaving them free to evolve and to manifest their own emergent properties through execution on a computer. Programming should be the preferred activity of philosophers because it allows them to test the models of the world they propose, provided that these models are unambiguous, complete, and finite \cite{butcher2012}. Programming is thus a creative action and a philosophical exercise; software, therefore, cannot be considered merely as a tool, but can by extension also be considered an \textit{art object}. Not only the output of a program, but the program itself, the \textit{very process}, can be art \cite{knuth-art, sep-computer-science, galanter2003generative}.

Software can, therefore, be situated within a broader framework: that of the \textit{Generative Arts} \cite{galanter2003generative}. As early as 1920, Naum Gabo wrote that a work of art is not an object, but a process: something that manifests in an object without, however, coinciding with it \cite{gabo1920}. The system is not a transparent medium for authorial intent, but the very body of the work itself.
\footnote{Although Galanter, in \cite{galanter2003generative}, defines the system as the \textit{method} through which the work is realized, he also recognizes cases where the system itself coincides with the artwork, demonstrating a degree of variability that will be revisited later in this paper.}

Consequently, the figure of the artist undergoes a profound transformation. Ceasing to be the \textit{Demiurge}, the craftsman of the finished object, they evolve instead into a \textit{systems designer}: Platonic in modeling the Hyperuranion from which a Demiurge will derive forms, and Aristotelian in defining the processes that will shape matter.

\section{Genealogy of Generative Creativity}
\label{sec:storia}

Once software has been established as a legitimate artistic outcome, it becomes possible to engage in a more specific discourse on Generative AI. To do so, it is useful to frame AI within the broader historical context of Generative Arts; not treating it as an anomaly, but rather reading it as an industrial-scale manifestation of a deeply rooted historical and cultural process. This discussion traces several stages of this process (among many others) that help frame the artist as a \textit{negentropic curator} (one who \textit{reduces} entropy) and an \textit{explorer} of latent spaces (understood both as specific spaces within the machine learning cotext and as generic spaces of the possibilities inherent in a system).

An essential stage of this journey is the analysis of fugues. A fugue is a musical form in which, starting from a small thematic fragment, an entire composition can be developed. This inherently possesses a generative nature, but there is one composer in particular worth mentioning. Around 1750, Johann Sebastian Bach undertook a monumental feat by studying all the fugues written up to that point. After studying this corpus, he synthesized \textit{The Art of Fugue} \cite{bach1750kunst}, which is regarded as one of the most profound representations of the concept of the fugue. What Bach achieved is not far removed from what occurs with AI: training a neural network (biological, in his case) to generate new works.

Also in the 18th century, the \textit{Musical Dice Game} emerged. This game consists of writing small musical fragments, associating them with numbers, and then using dice rolls to assemble these fragments into a single sequence. The quality of the result depends entirely on the quality, compatibility, and \textit{modularity} of the written fragments. This is not entirely different from what occurs when using Generative AI systems to create music. The difference lies in the fact that, in AI, the stochasticity of the process is heavily influenced by the training data and the prompt (though not always in a deterministic manner). However, the importance of the initial musical corpus remains central to both: even if elements are not simply copied and pasted but instead encoded within the neural network weights, the quality of the training data is a crucial factor for the quality of the output. A significant version of the game was formalized by Johann Kirnberger in 1757 \cite{kirnberger1757}, who worked extensively not only to select fragments but also to define some rules, leaving the player with the task of rolling the dice and judging the results.

Continuing the parallel between history and modern AI, one must consider László Moholy-Nagy's work \textit{Telephone Pictures} \cite{moholynagy1923}, a Bauhaus professor who in 1923 attempted to use telephone mediation as an artistic experiment. Moholy-Nagy calls an enamel company and dictates the work he intends to create over the phone. The worker, likely unaware of the artistic purpose, attempts to execute the instructions as best as possible, thereby creating the piece. This serves as an \textit{ante litteram} archetype of prompt engineering. Indeed, Moholy-Nagy's intention was to emphasize the role of the artist as a \textit{producer of concepts} rather than as a craftsman physically involved in the realization of the work.

Also in the twentieth century, Bruno Munari wrote \textit{Regola e Caso} (Rule and Chance) \cite{munari1992}, in which he essentially praises the balance between rule and randomness, between determinism and complete aleatoricness. He argues that determinism is boring, complete randomness is too unsettling, and that art must find the equilibrium between the two. Balancing determinism and randomness is indeed what all artists do, but it is also what neural networks do: there is a strong deterministic component emerging from the data, yet the final output is deliberately rendered slightly aleatoric via a \textit{temperature} parameter (cf. Sec.~\ref{sec:genai}) that ensures a certain degree of variability and unpredictability \cite{holtzman2020}.

It is worth noting that Munari had already radicalized this insight in his 1952 \textit{Manifesto del Macchinismo} (Manifesto of Machinism) \cite{munari1952}, in which he explicitly urged artists to ``distract machines by making them function irregularly.'' This vision foreshadows the practice of forcing systems out of their statistical equilibrium to achieve aesthetically interesting outcomes, and (alongside the discourse on aleatoric processes) anticipates Galanter's reflections on \textit{complex systems} (systems poised between perfect order and complete disorder) and their exploration \cite{galanter2003generative}.

When discussing aleatoricism, one cannot avoid mentioning John Cage, a twentieth-century composer who held a radical view regarding the importance of the composer. Cage rejected the cult of Western myth; according to him, the human being was not the author of a work: for Cage, the human being is a \textit{liberator of music} \cite{cage1961} that already exists in nature. Through his compositions, he sought to completely expel the concept of choice from the creative process, dismantling the European idea of music based on the centrality of the composer.

Cage's perspective, at first glance, seems to resonate with Alfred Kroeber's superorganicism \cite{kroeber1917}, which posits that the role of the individual is nearly negligible compared to the influence of society. Being shaped by a preceding culture, the act of creation becomes merely serving as a conduit for an idea already in the air; the individual acting as a manifestation and inevitable expression of the entire culture. In other words: had Mozart not been born, music similar to his would likely have emerged from another figure within that specific historical period and cultural milieu.

AI, with its massive volume of human training data, can be read as a technical and circumscribed manifestation of superorganism. However, there is a fundamental distinction between Cage and Kroeber worth addressing. Unlike dice, the I Ching, or any other aleatory device, society (and AI) does not operate \textit{only} by pure chance. Cage sought the silence of the ego; he aimed to nullify human intention to let nature speak. Modern language models do almost the opposite: they generate the white noise of thousands of human egos crystallized within the training data. Cage wanted art without memory; AI is, by definition, pure memory \cite{norouzi2026canvas}. Two things that appear similar on the surface but are profoundly antithetical, converging only in the idea of the dissolution of the \textit{single author}.

\subsection{Machines: A Taxonomy of Generative Systems}
\label{sec:macchine}

Another relevant example can be found in Jean Tinguely's \textit{Méta-matics} \cite{tinguely1959}. These are machines powered by an electric motor that, through nearly random movements dictated by the shape of rotating cams (knowledge crystallized into a deliberately imprecise mechanical ``software''), draw upon sheets of paper. They are \textit{machines that draw} or rather, \textit{works of art that draw}. Beyond the obvious parallel with image-generating AI, what warrants closer examination is the fact that Jean Tinguely distributed these machines not as instruments, but as works of art. He left the viewer to grapple with questions such as: ``Are the drawings it produces standalone works of art? Are they parts of the same work, like detached metastases that people can keep at home? Or are they merely a byproduct of the work?''

Continuing with the discussion of machines that create art, one must cite the work of the Barron spouses, who in 1956 created what is considered the archetype of science fiction film soundtracks. For the \textit{Forbidden Planet} soundtrack \cite{barron1956}, they created small electronic circuits that emitted sounds; they then recorded these sounds while \textit{torturing} the small electronic circuit, causing it to degenerate and self-destruct. Through careful editing of the recorded materials, they subsequently compiled the soundtrack. This is an example of machines creating art, used not as the work itself, but as a medium for creating a fixed work, specifically based on the \textit{aesthetics of failure} and destruction (anticipating the concept of \textit{Circuit Bending}).

Up to this point, computers have not yet been discussed. An early instance relevant to this study can be found between 1956 and 1957, when the chemist Lejaren Hiller decided to abandon chemistry to devote himself to computer-generated music. Together with Leonard Isaacson, Hiller created the \textit{Illiac Suite} \cite{hiller1957, hiller1959}, a four-movement composition considered the first machine-generated composition. They implemented software designed to compose an original score. The software combines sets of rules, statistical models, and mathematical models (thereby crystallizing knowledge across various modalities). This provides another example of a process that is not the artwork itself (or rather, is not marketed as such) but merely a medium through which a specific artwork was created (albeit one that is potentially reusable).

By the 1960s, it became evident that the artistic use of machines had become influential across all forms of human expression. The so-called \textit{3N} group (Frieder Nake, Georg Nees, and Michael Noll, considered a group more for chronological than geographical reasons) \cite{noll1967} began to popularize the idea of computer-based Generative Art, laying the foundations for modern \textit{creative coding}. At the same time, Nanni Balestrini was working on his \textit{Tristano} \cite{balestrini1966}: a literary work composed of fragments from various textual sources, randomly recombined and printed (at least in the author's original vision, which only materialized in 2007 \cite{balestrini2007}) as unique copies, each the result of different executions of combinatorial software. Both works are examples of \textit{process as artwork}, whose results (always different yet always similar) raise the same questions as Méta-Matics. These artifacts are neither originals nor copies: they suggest the necessity of a third ontological status (cf. Sec.~\ref{sec:noise}).

Returning to the realm of music, but (unlike the \textit{Illiac Suite}) citing a case where the process \textit{is} the work, one can discuss the work of Brian Eno and how he popularized the concept of generative music. To create \textit{Music for Airports} \cite{eno1978}, Eno designed loops of varying lengths to be played simultaneously, so that they interlock differently with each repetition, creating a harmonic texture that only repeats after very long intervals. Eno would have liked to deploy a system that allowed for the experience of this generative quasi-infinity (analogous to Balestrini, who intended for every copy to be unique), but he had to settle for a fixed duration of 48 minutes. The original artwork is meant to be experienced in its full combinatorial and mutable nature; listening to only one pre-recorded version is perceived as a rupture, betraying the fundamental centrality of the process that generates it.

Across all these examples, beyond the parallels with AI, three primary ways of interpreting generative systems emerge:

\begin{itemize}
    \item they can serve as \textit{media} for producing a specific, fixed work, such as the \textit{Illiac Suite};
    \item they can be parts of the entire \textit{artwork}, such as \textit{Music for Airports}, where the process is inseparable from the experience of the work;
    \item or they can function as \textit{instruments}, distributed as permanent generators of content, such as the \textit{Musical Dice Game} (which does not produce a single artwork, but rather serves as a tool to produce many).
\end{itemize}

A work does not unequivocally fall into one category or another \textit{a priori}. The distinction lies in the mode of engagement and publication, rather than in the system's structure. It is the artists (as well as critics and audiences) who determine which of these three categories the object under examination belongs to. This suggests that any general definition of generative art (e.g., \cite{galanter2003generative}) should move away from framing the system as either a tool or an end in itself, leaving such a classification open to each individual work.

This taxonomy applies as much to Generative Art in general as it does to Generative AI in particular. For example, the researchers behind ChatGPT, Suno, or Stable Diffusion could have conceived of them as \textit{works of art}; generative works that produce an infinite array of textual, musical, or graphic manifestations upon a user's request. Instead, they chose to market them as \textit{instruments}. Had they made the former choice, public perception of the Generative AI phenomenon might be very different.

\section{On Statistical Prediction}
\label{sec:genai}

It is useful now to briefly recall how AI works in order to highlight certain aspects relevant to the present discussion. As illustrated by XKCD comic strip no. 1838 \cite{munroe2017}, for a computer scientist, AI is a system capable of providing \textit{apparently} intelligent responses. They do not need to be truly intelligent; they only need to appear so. A Large Language Model, in particular, has no way of distinguishing truth from falsehood; it merely produces extremely plausible statements; it is up to the user (or an automated grounding system) to verify them.

A Neural Network (a specific type of machine learning system) is composed of input nodes that feed into \textit{hidden layers}. Numerical values flow between the nodes through weighted connections (the model's \textit{parameters}), ultimately producing an output. It is a simple structure from which, thanks to its enormous scale (and the presence of non-linear activation functions within the nodes), extraordinarily complex behavior emerges.

A Large Language Model (LLM) is a specific type of Neural Network capable of associating each word fragment (\textit{token}) with a series of numbers (\textit{embeddings}). The numerical values of these words are updated based on other words in the input (for example, through the \textit{Transformer} architecture and \textit{attention} mechanisms \cite{vaswani2017}) to predict the most probable next word in a context-dependent manner. For example, when chatting with such a system, the most likely word following the end of a question will be the beginning of the response.

An example: Consider the input

\begin{verbatim}
The quick brown fox jumps over the lazy
\end{verbatim}

This sentence, used to display fonts because it contains every letter of the alphabet, should be completed with \texttt{dog}. A trained LLM will have read it thousands of times and will almost certainly respond correctly. In reality, however, the model does not choose the most probable word: it samples randomly according to a probability distribution across a range of possibilities; a distribution regulated by a parameter called \textit{temperature} \cite{holtzman2020}. With very low temperature, the behavior is deterministic (always selecting the most probable word) but boring; by raising the temperature, other words gain the chance of being sampled: \texttt{cat}, \texttt{bunny}, \texttt{?} (Cf. \textit{Regola e caso} by Bruno Munari). This margin of randomness, this \textit{semantic dithering}, can cause a \textit{butterfly effect} capable of driving interactions with LLMs in unpredictable directions, and constitutes a sort of \textit{intentionality gap}.

What would happen if the model chose \texttt{cat} instead of \texttt{dog}? Word by word, seeking combinations statistically coherent with what has already been written, it could produce:

\begin{verbatim}
The quick brown fox jumps over the lazy
cat. I wrote "cat" because I was getting
bored with this absolutely overused
sentence.
\end{verbatim}

It seems as though the network possesses consciousness and a sense of humor, but in reality, it has merely stumbled due to high temperature, continuing to generate statistically plausible text from what it had already produced.

This is the behavior of a pure transformer, without the corrective systems and optimizations of modern models. But the underlying mechanism remains: most LLMs generate the next unit of information, feed everything it has generated back into the input, and repeat the process until the token indicating the end of the response is extracted.

Everything is based on syntax: on the frequency with which certain words appear near others. However, it is a syntactic statistics so vast and profound that it creates nearly the illusion of semantics\footnote{Within the framework of distributional semantics \cite{harris1954, firth1957}, the question remains as to whether semantics itself is reducible to large-scale syntactic structure. A literary example is provided by Roland Barthes in ``The Death of the Author'' \cite{barthes1967}, where he argues that writing is an impersonal process; one in which it is not the subject who speaks, but language itself expressing and articulating itself. Nevertheless, the core issue remains: an LLM lacks experiential knowledge of what it discusses.}. A neural network does not know that an apple can be red or green: it has only seen the words ``red'' and ``green'' near ``apple'' countless times, yet it possesses no concept of any of the three. So much has been said and written about the world that, by training networks on this vast amount of material through sheer brute force, they can even solve problems that ostensibly require logic and semantics. Yet occasionally (more often than one might think), syntactically perfect, highly plausible, yet false sentences emerge. This is the phenomenon of hallucinations, which betray the purely statistical nature of these systems \cite{ji2023} (along with other related issues, such as poor training or inadequate attention windows).

However, hallucinations can also be viewed as \textit{mutations}: unexpected deviations, secret passages into territories that no human being would have deliberately explored. A generative system is, in potential, an extraordinary \textit{Serendipity Machine} \cite{green2000} (much like Balestrini's \textit{Tristano}, or Brian Eno's \textit{Oblique Strategies} \cite{eno1975}, those cards used in recording studios to break creative deadlocks): even the statistical irregularity of a neural network (devoid of any intentionality) can open pathways that an artist would tend to dismiss \textit{a priori}. The condition is that someone must be ready to recognize their value.\footnote{It should be noted that this holds true only within the realm of art. In medical, legal, informational, or scientific contexts, hallucinations remain flaws to be minimized.}

The imagery of ``statistical blenders'' often used to describe these systems, while useful as a first approximation, runs the risk of being misleading and fails to capture their sophistication. It would, in fact, be more accurate to speak of \textit{statistical resonators}. This distinction is technically precise for architectures based on State-Space Models \cite{gu2023mamba}, whose mathematical definition essentially constitutes an adaptive filter bank\footnote{The eigenvalues of $A$ are the poles. $B$ and $C$ determine the zeros. $A$ determines the conceptual structures upon which the system resonates; $B$ decides which modes are excited by the input; $C$ decides which are observed to produce the output.}. As a text is read, resonances shift: the system tunes itself to certain linguistic structures. It does not \textit{blend}, but rather \textit{resonates}. For Transformer architectures \cite{vaswani2017}, the attention mechanism performs an analogous tuning, albeit without the same direct justification in terms of systems theory. The philosophical implications are significant: if the output were merely an interpolation, it would be a degraded copy of something already existing. Instead, it is something more specific: not a copy, nor an original, but the manifestation of the model's dynamic behavior excited by that context, subject to all constraints regarding filter order, as well as the fundamental limitation of filters: \textit{it cannot amplify a signal that does not exist}. One might object that in a non-linear context, the system is nonetheless capable of generating signals not present in the training set; however, nonlinearities can generate components not present in the input only as deterministic combinations of the incoming signals (harmonics, intermodulation products); the reachable space, however vast, remains a subset of the possible.

\section{Computational Creativity: Possibilities and Structural Limits}
\label{sec:novelty}

Generative models based on statistical learning tend to resonate with-, or gravitate toward-, the most heavily populated regions of their training space. The model is trained to produce the statistically most representative output for a given context \cite{manovich2018ai}: when asking Suno for a rock track, it produces something that is \textit{typically} rock; if asked for jazz, it delivers something that is \textit{typically} jazz; if asked for industrial noise experimental, it will produce something that is \textit{typically} industrial noise experimental (because at this point, it isn't even all that \textit{experimental}; these things have been done for fifty years now).\footnote{Anyone who feels threatened by generative AI music systems because they are encroaching upon ``their creative niche'' is likely operating within a creative space that is already extensively represented; conversely, those who create novelty (at least for now) feel more secure. This reminds of the ``head in the sand'' Turing Objection \cite{turing1950}, which in this context could be paraphrased as: The prospect of a machine being capable of creativity is so destabilizing that one finds it more reassuring to convince oneself that it cannot be. For further reading on the subject, see \cite{audry2021art}.}

This resonance with ``typical'' output is the default mode of Generative AI models, rather than an insurmountable boundary. Prompt engineering can be used to drive the model toward the less-traversed regions of its learning space, countering its tendency to respond with something reassuring instead of something that provokes aesthetic friction. The challenge for the artist becomes learning to navigate these low-density zones with awareness, and recognizing what occurs when the system begins to exhibit unexpected behaviors.

In this regard, it may be useful to introduce a theoretical distinction derived from studies on computational creativity. Margaret Boden distinguishes between \textit{Combinational Creativity} (the ability to recombine known elements to create new instances), \textit{Exploratory Creativity} (the ability to explore a given conceptual space, generating new instances while respecting its rules), and \textit{Transformational Creativity} (the ability to break the very rules of the conceptual space, inventing categories that did not previously exist) \cite{boden1990}. The latter is, at a higher level, a form of the second: exploring the space of possible conceptual spaces themselves \cite{wiggins2006}. AI tends to excel significantly in the former; it struggles more with the other two, which are structurally related. Though, the debate remains open: some recent models exhibit emergent behaviors that blur this boundary \cite{wei2022, schaeffer2023}. It is precisely where the model exhibits unexpected behaviors, straddling that very threshold, that the most interesting space for the artist resides.

But how can a deterministic system exhibit unexpected behaviors? This question originates in the notes of Ada Lovelace \cite{lovelace1843} and was later revisited by Turing \cite{turing1950}, who, however, provided an answer using generic arguments. To bolster Turing's argument more robustly, one could argue that it is a matter of perspective: the point is not determinism itself, but rather how practical it is to determine behavior within a useful timeframe. The emergent properties of a system are technically computable, yet the process is impractical (reminiscent of a well-known issue in cryptography, where the most secure encryption is that which cannot be broken within a useful timeframe).

It is worth adding an observation regarding the nature of aesthetic value in this context. Human aesthetic reception is largely independent of the intentions (or lack thereof) of the entity that materially generated a work \cite{wimsatt1946}. Beauty emerges in the eye of the beholder, not necessarily within the ``heart'' of the creator \cite{hume1757, jauss1982}. In computational aesthetics, what matters is not whether the machine experiences emotions, but whether it is capable of evoking them in the user. AI functions as a mirror to human expressive structures, trained to simulate their syntactic configurations; however, the machine's lack of emotion precludes it from being able to properly \textit{evaluate} according to human criteria (a specific form of alignment difficulty \cite{christiano2017, xu2023imagereward}).\footnote{Even assuming a form of embodiment for the artificial creative agent, such emotions would nonetheless not necessarily be attributable to human ones.} As Italo Calvino emphasizes in \textit{Cybernetics and Ghosts} \cite{calvino1967}, for syntactic generation (the \textit{cybernetics}) to transform into true art, human lived experience, choice, and responsibility (the \textit{ghosts}) remain irreplaceable \cite{Rohrmeier-2022}.

It is precisely these responsibilities that define the artist's role as a \textit{negentropic curator} and an \textit{explorer} of latent spaces. From the perspective of information thermodynamics, the models' latent space is a system of extremely high entropy. In this scenario, the artist ceases to be one who ``creates matter'' out of nothing, instead assuming the role (in the case of the \textit{negentropic curator}) of a modern Maxwell's Demon \cite{maxwell1871}: an entity that performs continuous selections at the cost of an expenditure of intelligence, separating the ``signal'' (meaning, emotional value, novelty) from the statistical background ``noise.'' There is, however, a subtle irony in all of this: for the neural network, ``noise'' is precisely what the training process sought to eliminate: statistical deviations, anomalies, and hallucinations. These are the very elements that, as we have seen, constitute the most fertile material for the artist. The artist's signal is the machine's noise, and vice versa; the artistic Maxwell's Demon works, so to speak, in a direction exactly opposite to the system it attempts to govern \cite{munari1952}. To abdicate this critical judgment would inevitably lead to the ``heat death'' of art.

\section{The Risks of Creative Automation}
\label{sec:risk}

The intensive use of generative tools exposes us to a highly insidious risk: a desensitization to \textit{cognitive effort}. This is a specific manifestation of the cognitive atrophy induced by chatbots \cite{kosmyna2025} (documented more broadly as \textit{automation complacency} \cite{parasuraman2010}); it does not involve a decline in faculties in the strict sense, but rather a progressive reduction in the tolerance for effort; a diminishing willingness to invest time and labor into a complex process \cite{risko2016} that is a necessary condition for any form of mastery. The musician generating music with AI, the writer composing with an LLM, or the programmer delegating code to an agent can easily achieve a result that lacks certain qualities \cite{anderson2024homog, padmakumar2024, perry2023}, qualities that are difficult to recover without friction with the process. The problem is that this result arrives in \textit{seconds} rather than \textit{days}, and the subject tends to settle for the available result \cite{risko2016}. Over time, what is missing may even cease to seem relevant \cite{glickman2024}. These tools can and should be used, but as \textit{tools}, not as \textit{delegations}. The final word (selection, correction, critical completion) must remain human; not as an ideological constraint, but because it is through friction with the process that the capacity to create (or even simply to judge) the result is maintained.

There is another phenomenon closely linked to this risk that is worth noting: Model Collapse \cite{shumailov2024}. It occurs when an AI model is trained on synthetic outputs generated by other models (of which the Internet is becoming increasingly saturated \cite{liang2024monitoring} due to so-called \textit{AI-Slop}). The system ends up resampling and degrading existing entropy in a sort of statistical inbreeding, losing all the variance that made it useful. The combined risk of Model Collapse and cognitive atrophy foreshadows a dystopian loop of progressive degeneration of creativity itself, both biological and artificial.

The antidote has a name: \textit{environmental enrichment}\cite{hebb1947}, borrowed from neurobiology and applied to the technological domain. For both its own well-being and that of the models, the human must be incentivized to actively inject three things into the system that, at present \cite{Rohrmeier-2022}, cannot be automated:

\begin{enumerate}
    \item extra-ordinary data, rare experiences, unprecedented configurations that the machine has never processed; everything that cannot be represented through the statistical distribution on which it was trained.
    \item embodied experience, Calvino's \textit{ghosts}, that which is rooted in the body and physical context, which no purely symbolic model can ever possess by definition \cite{damasio1999, varela1991}.
    \item critical judgment, which is the only operation that transforms a statistical generation into something meaningful \cite{manovich2018ai}.
\end{enumerate}

These are the things humanity must safeguard, yet they are also those at risk of being lost if we allow ourselves to be seduced by the convenience of the tool. Therefore, there is a call for policies that promote the production of human data: human creativity must be regarded as a common good to be protected.

Returning to the debate on \textit{Transformational Creativity}, Model Collapse demonstrates that there is indeed a limit to what a model can do. If AI were capable of true novelty, synthetic data would enrich the space; instead, it erodes it. AI, however sophisticated, operates within a manifold \cite{bengio2013} with degrees of freedom fixed by the training data: it can navigate it with extraordinary skill and extend it by pushing into its less frequented regions, but \textit{it cannot add a dimension that does not exist}. That capacity (to invent not just a new point in space, but a new axis, by operating at a meta-level \cite{Rohrmeier-2022}) remains, at least for now, the prerogative of biological agents.\footnote{However, even assuming that a future AI might be architecturally capable of producing new degrees of freedom, such elements would not necessarily be attributable to human experience, unless the new architecture replicates the human one \textit{exactly} (in both body and mind). For the modeling of the nervous systems of biological organisms, see the experiments in \cite{wangchen2024}.}

\section{Co-creativity and the Roles of Distributed Authorship}
\label{sec:cocreat}

The taxonomy of systems outlined in Sec.~\ref{sec:macchine} frames the status of the generative system; the taxonomy proposed here, instead, addresses \textit{who} contributes creatively, and at which stage.

To understand exactly where and how creativity and knowledge are injected into a Generative Art system, it is necessary to decompose it into a pipeline of primary components and processes:

\begin{enumerate}
    \item creation and selection of the data upon which the system is based (in terms of training or an explicit corpus used by the system),
    \item design of the architecture that defines its behavior,
    \item conditioning of the system through prompts (including multimodal ones) or other channels of interaction with the resulting system,
    \item the set of manipulations, but also curatorial and editorial choices applied to the produced material during \textit{post-generation}.
\end{enumerate}

What fuels each of these elements is always, albeit to varying degrees, \textit{human creativity}, and within each, one can recognize a variation of the roles that the artist is called to assume within the ecosystem of Generative Art.\footnote{Note that the proposed taxonomy focuses on the role of the \textit{artist} throughout the (potentially asynchronous) stages of the pipeline; it should therefore not be confused with similar taxonomies and classifications that focus on the role of the \textit{system} \cite{lubart2005can}, the interaction with it \cite{kantosalo2016modes, guzdial2019interaction, mixed-initiative}, or what it produces \cite{colton2011computational}.}

The idea that the artist's role is transformed by generative systems is not new. As early as 1973, Cornock and Edmonds described the artist as \textit{``a catalyst of creative activity''} within the context of computational art \cite{cornock1973creative}. This concept was later revisited by Norouzi and Prinz through the notion of \textit{catalytic collaboration} \cite{norouzi2026canvas}: a collaboration in which the machine is neither transparent nor predictable, and where the artist triggers processes that limit their direct control over the output (a limitation also expressed by \cite{galanter2003generative}). However, the present contribution rests on two different premises: 1.~the partner in the creative chain is not the algorithmic system (which, at present, remains medium, instrument, or artwork), but rather the collective of humans distributed along the Generative AI pipeline. 2.~the degree of control varies radically depending on the stage; this is not a generalized relinquishment, but rather forms of qualitatively distinct mastery, some of which (such as architectural design or post-production) involve fine-grained control that \textit{governs} the algorithm. The proposed taxonomy distributes these interventions across the four stages of the pipeline, assigning each its own name and specific creative responsibility.

\begin{description}  
    \item[Entropic agent] The first and most influential vector of creativity is the \textit{training corpus}. The contents comprising a model's training set (texts, images, scores, recordings) are largely the product of human beings, who have embedded their own worldview into shared culture. If every output resonates with and depends upon the totality of that corpus, then those who contributed to that corpus are co-authors in a non-metaphorical sense. This constitutes the most significant injection of creativity that a Generative AI system receives, yet it is also the most invisible: those who created the content are rarely aware they are contributing to the training of a machine. It is precisely here that the role of the artist as an \textit{entropic agent} takes root. The richness, originality, and embodied experience from which the data originates are not marginal details: they determine the quality and breadth of the entire latent space available to the system.\footnote{It is no coincidence that one of the most interesting trends in training personalized models is the use of small and highly curated datasets (even deliberately \textit{biased}), in which a narrow and coherent corpus is used to steer a model's behavior much more precisely and creatively than a vast, heterogeneous, and generalist dataset \cite{gal2023, ruiz2023}.} An interesting corollary to this point is that \textit{anyone}, the moment they create something, becomes a potential \textit{entropic agent} in service of future works.

    \item[Systems designer] The developers who have designed a model's architecture have, in turn, crystallized within it a set of non-neutral technical and aesthetic choices. The structure of a model is the encoding of a point of view on the world; it is not merely a technical act. As previously argued, it can be considered a philosophical and creative exercise. The model architect chooses which structures of the world to make representable, which relationships between concepts will become codifiable, and what form the space in which the system moves will take, along with its degrees of freedom. They are a \textit{systems designer}. Outside the context of Generative AI, this role may take on less computational contours (not all Generative Art is digital \cite{galanter2003generative}), but they remain essentially identical in substance.

    \item[Explorer] Then there is prompt engineering, often regarded as a second-order practice relative to the preceding stages. Yet, it involves its own form of mastery that should not be underestimated. Understanding a model (its space, its tendencies and blind spots, the emergent behaviors that characterize it) is a skill acquired through experience, leading to the ability to push the system toward regions where the unexpected resides. The prompt engineer explores a vast and partially obscured territory, learning to recognize its topography and to utilize its most peculiar recesses. They are, properly speaking, an \textit{explorer} of latent spaces. The same applies to generative art systems not based on AI: interactive multimedia installations that allow for user control can be experienced by the viewer as instruments through which to express intentionality, transforming the viewer into an element of the work or a co-author of it \cite{eco1962, ascott1990}. Conversely, works that do not involve forms of conditioning by the viewer concentrate the role of the explorer onto the systems designer, who also designs a mechanism for \textit{automatic exploration}.

    \item[Negentropic curator] Finally, there is the phase of \textit{post-generation} (as a generalization of ``post-production''): not merely selection \cite{colton2012}, retouching, or editing, but the entire act of \textit{framing}: the decision of what a work is, what it means, and how it is presented to the world. It is a fully realized creative act, known in the artistic field as \textit{curatorship} (a principle also suggested by Goodfellow \cite{goodfellow2024distributed}). As previously observed in Sec.~\ref{sec:macchine}, it is the agent governing the interface between the machine and the viewer who establishes the status of the system and its output. This governance is far from passive; it consists of choices that constitute creative acts in the fullest sense. Whoever operates in this phase performs exactly the function of the \textit{negentropic curator}: filtering the entropy of generative output through selection, correcting and refining results during post-production, and transforming a statistical distribution (or the entire system) into new meaning through curatorship.
\end{description}

But what happens when the artist does not manage the entire chain?  
Goodfellow proposes viewing the authorship of AI works as a spectrum, ranging from exclusively human creation to completely algorithmic \cite{goodfellow2024distributed}; here, however, the taxonomy of roles mapped onto the pipeline always refers to human authorship.\footnote{In this proposed view, Goodfellow's gradient can be interpreted as the \textit{degree of delegation} or the \textit{degree of plurality}. It could be interesting to measure the artist's contribution and the \textit{perceived weight} of each of the four stages of the pipeline, in order to quantify this gradation.}  
A generative system is the device through which, in fact, an asynchronous creative dialogue is realized with its designer and with everyone who has contributed (consciously or otherwise) to its creation. It is a collaboration that transcends spatio-temporal boundaries: when reread through the previously discussed examples, it reveals historical precedents of remarkable relevance. It is what Bach achieved by synthesizing the entire contrapuntal tradition in \textit{The Art of Fugue}; it is the logic of Kirnberger's Musical Dice Game, where he wrote the fragments and defined the rules, providing a system for players to roll dice and evaluate the outcomes; it is what Moholy-Nagy attempted by dictating instructions over Meucci's telephone to a worker unaware of the artistic purpose of their execution. In all these cases, collaboration occurred between individuals distant in space, time, and intention: the ``co-author'' was unaware of their role, yet they were performing it. Generative AI industrializes and radicalizes this pattern: the invisible co-author is no longer a worker in an enamel factory, but a multitude of human beings (contributing through the mediation of a neural network) whose presence in the creative chain remains, in most cases, entirely implicit.

In this light, the myth of the solitary ``romantic genius'' (already challenged by Cage through his radical dissolution of the creative ego, as well as by Barthes, Kroeber, Foucault, and many others \cite{barthes1967, kroeber1917, foucault1969}) definitively gives way to a concept of \textit{distributed authorship}. The work is no longer the fruit of a single individual, but rather the result of a chain of asynchronous collaboration that spans space and time, involving the original creators of the dataset, the architects of the model, those who formulate the prompt, and those who perform the final selection. Each of these stages incorporates a distinct and human creative act; the creativity of the output is an emergent property of the entire system (including humans). In this regard, the machine, in its current configuration, can only be a medium or a work in itself.

\subsection{Ethical and Regulatory Implications}
\label{sec:law}

This perspective, however, reveals an ethical issue that cannot be ignored. For collaboration to be legitimate, all parties involved must at least be aware of their involvement. Those who contributed training data have the right to be consulted, recognized, and compensated for this contribution \cite{gebru2021}. If this does not occur, the model of distributed co-creativity fractures: what could have been a collaboration transforms into unilateral appropriation, further compounded by widespread economic exploitation. A creative chain that fails to recognize all its links is not a form of collaboration; it is a form of exploitation \cite{gebru2021}. This is a real tension that has led to significant litigation and driven a necessary rethinking of the current regulatory framework.

On the regulatory front, several directions are emerging \cite{samuelson2023, euaiact2024}: the introduction of mandatory licensing for datasets used in training; the requirement to label generated content (``created with AI''); the extension of protections to voice, image, and personal style as forms of intellectual property; and copyright reform to address the large-scale generation of derivative content. This remains an ongoing process: the speed of technological progress has outpaced that of the regulatory response, as evidenced by the degree of impunity enjoyed (and, in part, continue to enjoy) by major industry players. In particular, the issue of liability for derivative content remains the least legally mature and the most ethically urgent. What is certain is that true distributed authorship requires, to be such, the explicit recognition of all its contributors.

\section{The Aesthetics of Failure: A Reevaluation of Old Neural Models}
\label{sec:oldmodels}

The Barron's practice of ``torturing circuits'' (which their contemporaries did not consider ``music'') is an example of the deliberate misuse (or tampering) of an instrument, driven by artistic inquiry and oriented toward an \textit{aesthetics of failure} \cite{menkman2011glitch}. In the recent history of Generative AI, these practices are explored by a very small niche relative to the rest of AI research \cite{sivertsen2024ambiguity}. Notable is Eugénie Desmedt's \textit{The Syntactic Synthesizer} \cite{desmedt2026}, an interactive installation that treats LLMs as analog instruments to be played (using physical knobs to control temperature, memory horizon, and semantic coherence). It is worth tracing a brief arc around this theme as well, because it is around this very concept that a potentially highly productive tension resides.

In 2015, Google conducted an experiment to understand which features were captured by individual neurons in an artificial neural network trained to distinguish faces and other patterns \cite{mordvintsev2015}. The problem, as is always the case when working with neural networks, is understanding what is actually being learned; which representations are consolidating within the hidden layers, in its latent space. The solution was to reverse the computational flow: rather than propagating information from input toward output, it was propagated backward, reconstructing the pixel configuration that would maximize the activation of a specific neuron. Neuron values thus became controllable parameters, knobs that could be adjusted to modify the characteristics of an input image. The result was \textit{DeepDream}: images in which the network, much like a computational pareidolia, amplifies everything it perceives as familiar, projecting its own internal representations onto external forms. A system trained to accurately recognize the world was forced (or \textit{distracted}, to use Munari's term \cite{munari1952}) to deform it according to its own internal logic.

What makes DeepDream artistically interesting is not merely its visual output (which is particularly lysergic), but the nature of the gesture that produces it. Forcing a neural network to manifest its own internal structures rather than describe the world is, structurally, the same operation performed by \textit{Circuit Benders}, and before them, the Barrons, on electronic devices: one does not work \textit{with} the tool, but \textit{against} it, in the zone of friction between intended function and the behavior emerging from abuse.

In this mode, latent space is conceived in a radically different way: it is not merely navigated, but altered by operating at a meta-level, modifying its architecture to exponentially extend its possibilities. In certain respects, it is transformed into a sort of Library of Babel \cite{borges1944}: an immense archive, largely dark and random, in which works that the system could not have generated spontaneously also exist in potentiality. The artist therefore assumes all roles simultaneously in this mode, including the traditional ones: \textit{craftsman of the object}, \textit{entropic agent}, \textit{explorer}, \textit{systems designer}, and \textit{negentropic curator}.

The problem is that (for understandable reasons) research has systematically drifted away from this territory. Starting from aesthetically \textit{radical} models (imperfect, unpredictable, and original precisely in their imperfection), much effort has been directed toward achieving something robust, realistic, and \textit{well-tempered}. Contemporary generative models are optimized to resist error: even when provided with an arbitrary input sequence, they will nonetheless produce a coherent, plausible output. This robustness, presented as progress, is also a loss: the model no longer produces anomalous outputs, no longer exhibits peculiar stereotypes, and no longer yields the unexpected. As technically extraordinary as it may be, it tends toward something we already know how to do: a photograph, a rendering, or a commissioned illustration. The constraint, paradoxically, stems not from the imperfection of the instrument, but from its perfection.

The older models should not be thought of as \textit{worse}, but as \textit{different}, characterized by a behavior that contemporary engineering seeks to avoid: the production of anomalies. Drawing upon the distinction between \textit{combinatorial} and \textit{transformational} creativity, these models prove paradoxically closer to the latter, not because they autonomously produce radical novelty, but because their specific mode of failure is a dimension (a degree of freedom) within which the artist can operate, and one that was not present before their advent.

\section{Noise, Power, and Aura}
\label{sec:noise}

In 1977, Jacques Attali proposed a radical thesis: music does not reflect society, but precedes it. Whoever controls sound controls social order. Attali identifies four stages in the relationship between sound and power (\textit{Sacrificing}, \textit{Representing}, \textit{Repeating}, \textit{Composing}) and describes modernity as the era in which music becomes a standardized and accumulable commodity, envisioning within the \textit{Composing} stage (in which anyone creates for the pure pleasure of doing so) the only possible subversion of the commodified order. \cite{attali1977}

In these pages, it has been shown how the tendency of modern AI to gravitate toward the center of the distribution, to excite its most represented modes behind a promise of \textit{Composition}, has actually accelerated an ongoing process of cultural homogenization (the publishing world was already saturated with \textit{Muzak} and \textit{Human-Slop} before generative models arrived), leading to what could be defined as \textit{Algorithmic Repetition}. However, a clarification is urgently needed: the issue is not \textit{AI yes or AI no}, nor is it \textit{old models vs. new models}, but rather \textit{how and with what intentions they are used}. The decline in diversity \cite{anderson2024homog} is not an intrinsic property of language models, but a specific effect of alignment processes (such as RLHF), of \textit{tuning}, that renders them safer and more predictable \cite{padmakumar2024}; it is up to the \textit{explorer} to push them into contexts far from their zone of statistical comfort.

Furthermore, it should be noted that the effects of \textit{Algorithmic Repetition} are also the causes of the Model Collapse discussed in Sec.~\ref{sec:risk}: both describe a system that resamples itself, reducing its own variance; the former on a cultural level, the latter on a statistical one. This also shifts the political issue raised by Attali onto a technical plane.

Attali's thesis also resonates with that of Walter Benjamin \cite{benjamin1935}: both diagnose what occurs when a work of art enters the era of its technical reproducibility. For Attali, it is a matter of power; for Benjamin, it is the loss of aura (that dimension of uniqueness that mass reproduction would dissolve). In a generative context, however, the picture becomes interestingly complicated: by definition, \textit{every generative output is unique} (at least within the assumed operating regime). According to Benjamin, therefore, the aura should not dissolve; rather, it condenses upon the generative system that produces the artifact \cite{fernandez2023ai}. Aura as a \textit{localized excitation emerging directly from the cultural field} that gave rise to the model. This shift in the aura requires a third term to describe the output of a generative system, one that transcends the traditional opposition between \textit{original} and \textit{copy}, which is proposed here as \textit{manifestation}.\footnote{Used here in its ontological sense (cf. Gabo, Sec.~\ref{sec:coding}), not in the sense popularized by contemporary ``manifesting'' culture.}

For example: a unique object produced by a generative system (whether it be a drawing made by Méta-matics, a drawing created by Nees's software, or a piece produced by an AI) is not an \textit{original} because infinite other equivalent instances exist, nor is it a \textit{copy} because it is not the reproduction of a pre-existing object: it is a \textit{manifestation} of a process, an \textit{alias} for an aura residing elsewhere. The \textit{manifestation} is thus the new status of the artwork in the era of generative production: a unique, unrepeatable object that is simultaneously non-individual. What can be endowed with the adjective \textit{original}, imbued with aura, is the process that generated it.

Within this framework, it is useful to revisit the distinction proposed by Nelson Goodman \cite{goodman1968languages} between \textit{autographic} and \textit{allographic} arts. An autographic work is one in which even a faithful replica remains a forgery: authenticity depends on the object having been produced by a specific hand at a particular moment (such as a painting). In contrast, an allographic work is identified by its adherence to a notation: any correct performance of a score \textit{is} the work itself, not a copy of it. Music, for example, is allographic: there are no forgeries of Bach; there are only performances that either conform to or deviate from what Bach wrote.

Within this topology, a generative process can be defined as allographic: for instance, code is an executable notation, and its outputs are executions. However, the analogy with the musical score is short-lived: the correct performances of a symphony are notationally equivalent, whereas the executions of a generative system produce objects that are unique each time\footnote{Far more variable than any performance of written music could be. One might object that Terry Riley's \textit{In C} \cite{riley1964} is a work composed of ever-changing performances; however, it can readily be classified as a generative work, and thus such performances are best described as \textit{manifestations} of the work.}. Yet, generative output is neither autographic (lacking a production history that confers uniqueness) nor strictly allographic (as it is not interchangeable with other outputs from the same system). It escapes both categories; for this reason, too, the term \textit{manifestation} is proposed as a third ontological status; the \textit{manifestation} of a process, endowed with singularity without autography, and with processuality lacking notational identity.

It should be noted that, in the same essay, Attali speaks primarily of \textit{noise}: for him, noise possesses no distinct technical or social significance; rather, it is inherently political, as established by the very definition provided by information theory. Attali demonstrates that what a society identifies as disturbance systematically coincides with those voices that threaten the codified order: minorities, unaligned aesthetics, and aspects of reality intended for marginalization. Contemporary models, trained on mainstream corpora and optimized for statistical robustness, replicate this exact gesture: they amplify the center of the distribution and marginalize anything that deviates from it, acting, without explicit declaration, as apparatuses of power.

It is therefore argued that older models (or, more broadly, those systems that manifest and exploit, more or less deliberately, the noise of a culture) are indeed bearers of an aura: they generate authentic difference in the sense described by Attali, rendering the process recognizable, irreplaceable, and \textit{original}. In contrast, modern frontier models (with their forced robustness and pronounced convergence toward the center of dominant cultures, even if they do not generate \textit{copies}) are themselves, at a higher level, \textit{copies} of a cultural process (the technical reproduction of something that already exists). If culture itself is instead considered a generative process, then these models are its \textit{manifestation}, just as the output is that of the model. Both interpretations converge on the same conclusion: in its naive use, the frontier model introduces no inherent difference relative to the cultural field from which it originates; it thus remains devoid of an aura distinguishable from that of the culture that produced it. Thus, \textit{Algorithmic Repetition} operates in a cascade across two levels: that of the model relative to culture, and that of the output relative to the model.

These models can only succeed in claiming an aura of their own if artists manage to do with them what the Barrons did with circuits, and what Munari theorized (perhaps provocatively) in the \textit{Manifesto del Macchinismo}: to use them as an object to be tortured, forced, and bent to one's will, so that their own voice \textit{and} that of the machine, along with its hidden imperfections, may emerge.

Therefore, if one wishes to leverage these new models in this manner, the artist must first act as an \textit{entropic agent}, dismantling the system to recover entropy where homogenization has artificially suppressed it, and subsequently act as a \textit{negentropic curator} to re-establish a new order. In this way, rather than being a device of power that amplifies the center of the distribution and perpetuates \textit{Algorithmic Repetition}, the generative system becomes a device of liberation that amplifies marginalized voices, and the artist becomes an agent of social change operating through the manipulation of noise.

\section{Conclusion}

In summary, drawing upon a historical genealogy and technological contextualization, this paper introduces: a taxonomy of functional categories for generative systems (\textit{medium}, \textit{artwork}, \textit{instrument}), where attribution is an \textit{editorial act} rather than an ontological property; a taxonomy of four artist roles (\textit{entropic agent}, \textit{systems designer}, \textit{explorer}, and \textit{negentropic curator}) based on the premise that the creative partner in the generative chain is not the algorithmic system (which, for now, remains either medium, instrument, or artwork), but the collective of humans distributed throughout the pipeline, thereby establishing a framework of distributed authorship; \textit{environmental enrichment} as a response to Model Collapse and cognitive atrophy; \textit{Algorithmic Repetition} as an aesthetic degeneration of aligned generative systems; and \textit{manifestation} as a third status transcending the original/copy dichotomy.

Before concluding, from a perspective of retrospective positioning and comparison with the literature, it is worth verifying the robustness of the proposed contributions against Galanter’s nine problems for Generative Art, regardless of technology \cite{galanter2019artificial}. While these problems are radicalized rather than resolved by Generative AI, the work addresses at least seven of them. The \textit{problem of authorship} and the \textit{problem of intention} are addressed through the taxonomy of roles and the concept of asynchronous co-authorship (Sec.~\ref{sec:cocreat}). The \textit{problem of uniqueness} (the paradox of ``unique objects produced in mass'') finds an ontological resolution in the concept of \textit{manifestation}, which names what Galanter left as an aporia (Sec.~\ref{sec:macchine} and \ref{sec:noise}). The \textit{problem of authenticity} is addressed through the distinction between aesthetic value and generative intention, in continuity with Calvino (Sec.~\ref{sec:novelty}). The \textit{problem of dynamics} (whether art resides in the process or the artifact) is shown to be an editorial rather than ontological distinction within the taxonomy of generative systems (Sec.~\ref{sec:macchine}). The \textit{problem of locality, code, and malleability} (where the work resides—in the code, the system, or the output—and who holds the power to define it) is addressed in Secs.~\ref{sec:coding} and \ref{sec:macchine} and spans the entire contribution up to the concept of \textit{manifestation} in Sec.~\ref{sec:noise}. The \textit{problem of creativity} is developed through the distinction between different types of creativity applied to generative systems, and through the critique of \textit{Algorithmic Repetition} as its degeneration (Sec.~\ref{sec:novelty} and \ref{sec:noise}). Finally, the \textit{problem of postmodernity} finds its answer not in the dissolution of the referent (the \textit{simulacrum} \cite{baudrillard1981simulacres}), but in its displacement from object singularity to system processuality (Sec.~\ref{sec:noise}).

Finally, several social aspects must be reaffirmed: First, risks like cognitive atrophy and Model Collapse can be mitigated through \textit{environmental enrichment},namely policies safeguarding and incentivizing human creative activities. Another aspect involves a formalization of the artist's roles, which can be identified throughout the generative pipeline: \textit{entropic agent} within the data, \textit{systems designer} in the architecture, \textit{explorer} via prompting or interaction, and \textit{negentropic curator} in post-generation. Third, the legitimacy of distributed authorship (an asynchronous, largely unconscious co-authorship) requires explicit recognition for all contributors, including those providing training data in the case of Generative AI. The regulatory dimension, though moving in the directions discussed in Section~\ref{sec:law} (mandatory licensing, labeling, protection of voice and style, copyright reform), remains structurally lagging behind technological evolution and constitutes the primary area of ongoing development.

Finally, the last social aspect concerns the relationship between generative models and power: these become instruments of liberation only when ceasing an \textit{Algorithmic Repetition}, that is, when they stop gravitating around the center of a statistical distribution that marginalizes noise and the unexpected (the native behavior of imperfect models). In this sense, the artist's role is that of an \textit{entropic agent} and \textit{explorer} who dismantles the homogenization, and a \textit{negentropic curator} who restores voice to marginalized noise.

\section*{Methodological Note and AI Statement}

This article is a manifestation of the 
``Artificial Intelligence for Music''
course taught by the author at the University 
of Milan. 
Based on a talk transcript and a series of notes, the structure was organized into a coherent sequence through a clustering operation \cite{ward1963, murtagh2014} based on embeddings \cite{chen2024bge} and subsequent optimal linearization \cite{barjoseph2001}. Some of the resulting passages were merged and reworked using Claude Sonnet 4.6 (Anthropic) and Gemma 4 (Google DeepMind); the result was curated and rewritten for stylistic uniformity, semantic refinement, and bibliographic integration.\footnote{The process exemplifies the proposed taxonomy: the author as \textit{entropic agent} in material selection, \textit{systems designer} in clustering architecture, \textit{explorer} in prompt formulation for language models, and \textit{negentropic curator} in final rewriting.}

As for the models: behind 'Claude' and 'Gemma' lie vast, often uncredited teams, but above all, the most anonymous contributors: the authors of the corpus of millions of texts on which these models were trained.

\bibliography{Bibliography}

\end{document}